\documentclass[useAMS]{mn2e}
\usepackage{lscape,graphicx}

\newcommand{\ewr}{\mbox{$W_r(2796)$}}
\newcommand{\apg}{\:^{>}_{\sim}\:}
\newcommand{\apl}{\:^{<}_{\sim}\:}

\newcommand{\etal}{et al.}

\newcommand{\lya}{\mbox{${\rm Ly}\alpha$}}

\begin{document}

%\slugcomment{Manuscript for The Astrophysical Journal}

\title[The $W_{\rm Mg\,II}$--$L_{\rm [O\,II]}$ Correlation in SDSS QSO Spectra]{On the Observed $W_{\rm Mg\,II}$--$L_{\rm [O\,II]}$ Correlation in SDSS QSO Spectra}
%\title{On the Observed [O\,II] Luminosity and Mg\,II  Absorption in
%  SDSS Fibers}
\author[L\'opez \& Chen]{Gilberto ~L\'opez$^{1}$
%\thanks{E-mail:glopez@fas.harvard.edu} 
and Hsiao-Wen Chen$^{2}$\thanks{E-mail:hchen@oddjob.uchicago.edu} \\
\\
$^{1}$Department of Astronomy, Harvard University, Cambridge, MA 02138 \\
$^{2}$Department of Astronomy \& Astrophysics, and Kavli Institute for Cosmological Physics, University of Chicago, Chicago IL 60637 }

%\pagerange{\pageref{firstpage}--\pageref{lastpage}} \pubyear{2011}

\maketitle

%\label{firstpage}

\begin{abstract}

  This paper investigates the effect of differential aperture loss
  with SDSS fibers and examines whether such selection bias would
  result in the observed correlation between rest-frame absorption
  equivalent width of Mg\,II absorbers, $\ewr$, and mean associated
  [O\,II] luminosity, $L_{\rm [O\,II]}$, in SDSS QSO spectra.  We
  demonstrate based on a Monte Carlo simulation that the observed
  \ewr\ vs. $L_{\rm [O\,II]}$ correlation of Mg\,II absorbers can be
  well-reproduced, if all galaxies found in deep surveys possess
  extended Mg\,II halos and if the extent of Mg\,II halos scales
  proportionally with galaxy mass as shown in previous studies.  The
  observed correlation can be explained by a combination of (1) the
  known \ewr\ vs. $\rho$ anti-correlation in galaxy and Mg\,II
  absorber pairs and (2) an increasing aperture loss in the $3''$
  diameter SDSS fiber for galaxies at larger $\rho$.  Galaxies at
  larger projected distances produce on average weaker Mg\,II
  absorbers and weaker (or zero) $L_{\rm [O\,II]}$ in SDSS QSO
  spectra.  We show that such correlation diminishes when larger
  fibers are adopted and is therefore not physical.  While under a
  simple halo model the majority of Mg\,II absorbers do not directly
  probe star-forming disks, they trace photo-ionized halo gas
  associated with galaxies.  We show that because of the scaling
  relation between extended gas cross-section and galaxy mass, the
  number density evolution of the Mg\,II absorber population as a
  whole provides a good measure of the cosmic star formation history.
%  A revised formula for
%  estimating the comoving star formation rate density based on
%  observed absorber number densities is provided in this paper.
%  Because models that do not require outflows can also reproduce the
%  empirical correlations between absorber abundances and star
%  formation properties, we caution drawing conclusions in favor of an
%  outflow origin for QSO absorbers based on these simple correlations.
%  Finally irrespective of the physical origin of the \ewr\ vs.\
%  $L_{\rm [O\,II]}$ correlation, such observed correlation offers a
%  convenient tool for estimating the comoving star formation rate
%  density based on the observed number density of Mg\,II absorbers.

%  Adopting a simple model for describing gaseous halos around galaxies
%  and empirical relations for describing the luminosity and size
%  distributions of the general galaxy population, 

%  Mg\,II absorbers are routinely found in the vicinities of galaxies,
%  and galaxies at larger projected distances $\rho$ to a QSO sightline
%  are found to produce progressively weaker Mg\,II absorbers in the
%  QSO spectrum.  

\end{abstract}


\section{INTRODUCTION}

Absorption line spectroscopy is a powerful tool for studying the
structure of the distant universe.  By observing the absorption
features imprinted in the spectra of background QSOs, we can study
otherwise invisible gaseous structures to redshift as high as
background QSOs can be.  The Sloan Digital Sky Survey (SDSS; York
\etal\ 2000) has yielded optical spectra of $\sim 106$\,k QSOs at
$z=0.065-5.46$ (Schneider \etal\ 2010).  This unprecedentedly large
sample of QSO spectra offers a rich resource for studying the distant
universe using absorption spectroscopy.  For example,
Mg\,II\,$\lambda\lambda$ 2796,2803 absorption doublets are commonly
seen in QSO spectra obtained using ground-based spectrographs.  They
provide a uniform probe of intervening gas over a broad redshift range
from $z=0.4$ to $z\approx 2.5$.  Several groups have conducted
systematic searches of Mg\,II absorption features in SDSS QSO spectra,
producing a large sample of these absorbers for constraining their
statistical properties (e.g.\ Bouch\'e \etal\ 2004; Nestor \etal\
2005; Prochter \etal\ 2006; York \etal\ 2006; Quider \etal\ 2011).

Mg\,II absorbers are routinely found in the vicinities ($\apl 100$
kpc) of distant galaxies and provide a convenient probe of galactic
halos at high redshifts (e.g.\ Bergeron 1986; Steidel \etal\ 1994;
Kacprzak \etal\ 2008; Chen \& Tinker 2008; Gauthier \& Chen 2011).
However, their physical origin, whether the absorbers arise in
infalling clouds, outflows from nearby star-forming regions, or a
combination thereof, is unclear.  The ubiquitous presence of
blueshifted Mg\,II absorption features along the sightlines into the
star-forming regions of $z\sim 1$ galaxies indicates that starburst
driven outflows are common in distant galaxies and that outflows
contribute to some fraction of Mg\,II absorbers uncovered along random
QSO sightlines (e.g.\ Weiner \etal\ 2009; Rubin \etal\ 2010).
Although the extent of galactic-scale winds around these galaxies is
unknown, the observed self-absorption of Mg\,II doublets in distant
star-forming galaxies has motivated recent works that attribute all
strong Mg\,II absorbers (rest-frame absorption equivalent width $\ewr>
0.5$ \AA) along QSO sightlines to outflows (e.g.\ M\'enard \&
Chelouche 2009; Chelouche \& Bowen 2010; M\'enard \etal\ 2011).

%Previous studies have shown that the observed strengths of Mg\,II
%absorbers correlate strongly with galaxy mass, and weakly with optical
%colors or recent star formation history (e.g.\ Steidel \etal\ 1994;
%Chen \etal\ 2010a,b).  In addition, luminous red galaxies have been
%found to contribute to strong Mg\,II absorbers along QSO sightlines
%(e.g.\ Gauthier \etal\ 2010; Bowen \& Chelouche 2011; Gauthier \& Chen
%2011).  The findings of strong Mg\,II absorbers around quiescent
%galaxies indicate the presence of gas accretion in these massive
%halos.  Together, these empirical results show that the nature of
%Mg\,II absorbing gas is more complex than a simple outflows origin.

Using a sample of 8500 Mg\,II absorbers at $z=0.4-1.4$ from SDSS DR4
quasar spectra, M\'enard \etal\ (2011; hereafter M11) observed a
strong correlation between $W_r(2796)$ and their associated {\it
  median} [O\,II] luminosity per unit area, $\sum_{L_{\rm [O\,II]}}$,
in stacked QSO spectra.  The observed \ewr\ vs.\ $\sum_{L_{\rm
    [O\,II]}}$ correlation is characterzed by
\begin{equation}
\left\langle\sum\,_{L_{\rm [O\,II]}}\right\rangle_{\rm med}=a\left[\frac{\ewr}{1\,{\rm \AA}}\right]^b
\end{equation}
where $a=(1.48\pm 0.18)\times 10^{37}\,{\rm erg}\,{\rm s}^{-1}\,{\rm
  kpc}^2$ and $b=1.75\pm 0.11$.  This observed correlation applies to
Mg\,II absorbers of $\ewr=0.7-6$ \AA.  A similar trend has also been
mentioned in Noterdaeme \etal\ (2010), but these authors did not
find a correlation between $W_r(2796)$ and $L_{\rm [O\,II]}$ in
[O\,II]-emission selected Mg\,II absorbers.  In addition, Noterdaeme
\etal\ (2010) pointed out that part of their strong Mg\,II absorbers
arise in low $L_{\rm [O\,II]}$ galaxies.  Because [O\,II] luminosity
$L_{\rm [O\,II]}$ provides a measure of current star formation rate
(e.g.\ Kennicutt 1998), M11 interpreted the observed strong
correlation as Mg\,II absorbers tracing on-going star formation.  In
addition, M11 showed that the frequency distribution function of
Mg\,II absorbers and the [OII] luminosity function share similar shape
and amplitude, and that the number density evolution of MgII absorbers
follows the cosmic star formation history.  Combining these empirical
correlations, the authors argue that outflows are the mechanism
responsible for the observed Mg\,II absorption in QSO spectra.

The conclusion of outflows being responsible for the observed Mg\,II
absorbers appears to be discrepant from previous findings that \ewr\
correlates more strongly with galaxy mass and weakly with galaxy
colors or recent star formation history (e.g.\ Steidel \etal\ 1994;
Chen \etal\ 2010a,b).  Such conclusion also makes it difficult to
understand the identifications of strong Mg\,II absorbers in the
vicinities of quiescent galaxies (e.g.\ Gauthier \etal\ 2010; Bowen \&
Chelouche 2011; Gauthier \& Chen 2011).

While M11 presented a clever approach to estimate the co-moving star
formation rate density $\dot{\rho}_*$ based on the observations of
Mg\,II absorbers in SDSS QSO spectra, it is important to understand
the underlying factors that shape the observed strong \ewr\ vs.\
$\sum_{L_{\rm [O\,II]}}$ correlation in SDSS data.  We note that the
SDSS fibers have a finite size of $3''$ diameter on the sky, which
corresponds to projected physical separations of $11-18\ h^{-1}$ kpc
at $z=0.4-1.4$.  The observed $\ewr$ vs. $\rho$ anti-correlation
(e.g.\ Chen \etal\ 2010a) implies that galaxies at larger projected
distances produce on average weaker Mg\,II absorbers and lower (or
zero) $L_{\rm [O\,II]}$ (when the star-forming disks occur at angular
distances $\theta>1.5''$, or $\rho>6-9\ h^{-1}$ kpc, of the QSO
sightlines; see Figure 1).  This selection bias strenthens the
apparent correlation of Equation (1)\footnote{M\'enard \etal\ (2011)
  applied a fiber selection correction by considering luminosity
  surface density, $\sum_{L_{\rm [O\,II]}}=L_{\rm [O\,II]}(z) / A_{\rm
    fiber}(z)$, where $A_{\rm fiber}(z)$ represents the corresponding
  physical area of SDSS fiber at redshift $z$.  However, as
  illustrated in Figure 1, the reduction from total observed fluxes to
  flux surface density does not correct for the differential aperture
  loss of galaxy fluxes in SDSS QSO spectra.}.  A similar point on the
potential missed galaxy light in the SDSS fibers has also been made in
Noterdaeme \etal\ (2010).
%It appears that the observed \ewr\ vs.\ $L_{\rm
%  [O\,II]}$ correlation may be a result of differential aperture loss
%of SDSS fibers.

\begin{figure}
\includegraphics[scale=0.25]{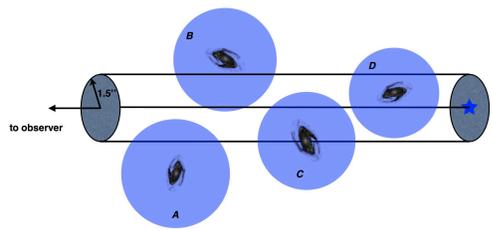}
\caption{Cartoon illustrating the fiber selection bias of intervening
  galaxies in SDSS QSO spectra.  Intervening galaxies at different
  projected distances to the QSO sightline may imprint absorption
  and/or emission features in the spectrum of the background QSO.
  Galaxies with optical disks overlapping a SDSS QSO fiber are
  expected to imprint [O\,II] emission features in the QSO spectrum.
  Galaxies with gaseous halos intercepting the QSO sightline are
  expected to imprint Mg\,II absorption features in the QSO spectrum.
  In this picture, the optical disks of galaxies $C$ and $D$ are
  covered by the fiber (in part or in full), galaxies $A$ and $B$ are
  not.  Only $C$ and $D$ are expected to contribute to the [O\,II]
  emission line seen in the stacked QSO spectrum.  On the other hand,
  the halos of galaxies $B$, $C$, and $D$ all intercept the QSO
  sightline, while $A$ does not.  Only $B$, $C$, and $D$ are therefore
  expected to contribute to the Mg\,II absorption statistics.  But
  because the QSO sightline impacts the outer part of galactic halo
  $B$, the pathlength through the halo is shorter and the Mg\,II
  absorbing strength is expected to be weaker (e.g.\ Chen \etal\
  2010a).}
\end{figure}

To investigate the selection bias as a result of differential aperture
loss, we have carried out a Monte Carlo simulation study.  Adopting a
empirical model for describing gaseous clouds around galaxies from
Chen \etal\ (2010a) and empirical relations for describing the
luminosity and size distributions of the general galaxy population, we
demonstrate in this paper that the observed \ewr\ vs. $L_{\rm
  [O\,II]}$ correlation can be well re-produced without any
fine-tuning after accounting for the differential aperture loss of
galaxy fluxes in the SDSS fiber.  While under the simple halo model
the majority of Mg\,II absorbers do not directly probe star-forming
disks like high column density damped \lya\ absorbers (e.g.\ Wild
\etal\ 2007), they do probe photo-ionized clouds around distant
galaxies\footnote{Note that the empirical \ewr\ vs. $\rho$ relation of
  Chen \etal\ (2010a) was established based on a sample of galaxies at
  close projected distances to a QSO sightline, including those that
  may produce a damped \lya\ absorption feature in the QSO spectrum.
  We therefore expect that absorber samples generated based on the
  mean \ewr--$\rho$ relation and the observed scatter include
  contributions from star-forming disks.  Based on the observed number
  densities of damped \lya\ absorbers (Rao \etal\ 2006) and Mg\,II
  absorbers of $\ewr\ > 0.5$ \AA\ (Nestor \etal\ 2005) at $z<1$, we
  estimate that no more than 20\% of these strong Mg\,II absorbers
  have contributions from star-forming disks}
%The success in finding damped \lya\
%absorbers in strong ($\ewr > 0.5$ \AA) Mg\,II absorbers (e.g.\ Rao
%\etal\ 2006) can be explained by an impact parameter selection.
%Galaxies at smaller impact parameters are expected to produce stronger
%Mg\,II absorbers, and the background QSOs have a higher probability of
%intercepting the star-forming disks and showing a damped \lya\
%absorbing feature (e.g.\ Rao \etal\ 2006).  
If more massive galaxies are surrounded by more extended Mg\,II
absorbing gas, then the number density evolution of MgII absorbers
naturally follows the cosmic star formation history.  Because models
which do not require outflows can also reproduce the empirical
correlations between absorber abundances and star formation
properties, we caution drawing conclusions in favor of an outflow
origin for QSO absorbers based on these simple correlations.
% We describe the process by which we generate th mock catalog of
% galaxy--Mg\,II absorber pairs in Section 2.  In section we present
% the catalog of galaxy-Mg\,II absorber pairs generated by our
% simulation.  Finally, we examine and discuss the results of our
% simulation and the observed correlation between [O\,II] luminosity
% and Mg\,II absorption in section .
Throughout this paper, we adopt a $\Lambda$CDM cosmology with
$\Omega_M=0.3$ and $\Omega_{\Lambda}=0.7$ and a dimensionless Hubble
constant of $h=H_0/(100\,{\rm km}\,{\rm s}^{-1}\,{\rm Mpc}^{-1})$.

%%%%%%%%%%%%%%%%%%%%%%%%%%%%%
%%%%%%%%%%%%%%%%%%%%%%%%%%%%%
%     Generating Mock Galaxy Catalog     %
%%%%%%%%%%%%%%%%%%%%%%%%%%%%%
%%%%%%%%%%%%%%%%%%%%%%%%%%%%%
%\clearpage
\section{Mock Galaxy and Absorber Catalogs}

To simulate the SDSS observations, we first generate a mock catalog of
200,000 galaxies distributed uniformly within $\approx 10''$ angular
radius of a QSO sightline at redshift between $z=0.4$ and $z=1.4$.
The large number is necessary to provide a representative sampling of
a wide range of galaxy properties (such as size and luminosity) and to
minimize statistical noise in our simulations.  The redshift range is
selected to match the study of M11.  Over the redshift range of
$z=0.4-1.4$, the angular radius of $10''$ corresponds roughly to
$\rho=37-59\ h^{-1}$ kpc.

The working hypothesis of this exercise is that all galaxies are
surrounded by extended gaseous halos.  This hypothesis is supported by
empirical observations that reveal a high gas covering fraction around
galaxies of a wide range of luminosity and color (e.g.\ Chen \etal\
2010a).  The gaseous halos are expected to produce Mg\,II absorption
features in the spectrum of a background QSO when intercepting the QSO
sightline.  For every Mg\,II absorber in the mock catalog, the
location and intrinsic properties of the absorbing galaxy are known.
It is therefore possible to compute the emission fluxes of the
absorbing galaxies recorded in QSO spectra.

\subsection{The Mock Galaxy Catalog}

We first generate random galaxies following a Schechter probability
function of their rest-frame absolute $B$-band magnitude $M_B$,
\begin{equation}
p(M_B)\propto10^{0.4\,(M_*-M)(1+\alpha)}\,\exp(-10^{0.4\,(M_*-M)}),
%\frac{L}{L^*}\right)^{-\alpha}\exp\left(-\frac{L}{L^*},
\end{equation}
where $\alpha$ is the faint-end slope of the galaxy luminosity
function.  We adopt $\alpha=-1.3$ for the faint-end slope (e.g.\ Faber
\etal\ 2007) and $M_{B_*}-5\,\log\,h=-19.8$ following Chen \etal\
(2010a).  The mock galaxy sample spans a lumionsity range between
$0.01\,L_*$ and $10\,L_*$.  As discusse below, we also repeat the
Monte Carlo simulations with different faint-end slop values,
$\alpha=-1.2$ and $\alpha=-1.4$ and our findings remain the same.

Next, we determine the optical size of each galaxy in the mock catalog
according to the luminosity-size relation from Cameron \& Driver
(2007),
% Luminosity-size relation
\begin{equation}
\log\left(R_{50}\right)= -0.1\,M_B-1.35
\end{equation}
where $R_{50}$ is the half-light radius, the radius within which the
galaxy emits half of its total flux.  For a galaxy of $M_B$, the
half-light radius is drawn from a Gaussian distribution about the mean
luminosity-size relation with a 1-$\sigma$ width of 0.4 dex in the
$\log\,R_{50}$ space.  To include the majority of the galaxy light,
the radius that contains 90\% of the total flux $R_{90}$ is more
relevant in our study.  We therefore convert $R_{50}$ to $R_{90}$,
assuming an exponential surface brightness profile.

Next, we determine the [O\,II] luminosity $L_{\rm [O\,II]}$ for each
galaxy assuming that the rest-frame $B$-band fluxes trace the mean
profile of [O\,II] emission.  The expected $L_{\rm [O\,II]}$ is then
calculated according to the correlation between $M_B$ and $L_{\rm
  [O\,II]}$ found in deep survey data.  Based on the study of Zhu
\etal\ (2007), we find that the $M_B$ vs.\ $\log\,L_{\rm [O \,II]}$
correlation can be characterized by
\begin{equation}
\log\,[L_{\rm [O \,II]}/({\rm erg}\,{\rm s}^{-1})]=35-0.3\,M_B
%M_B=-3.215\ \log(L_{[O \,II]})+109.275\,.
\end{equation}
with a 1-$\sigma$ scatter of $\sigma_{\log\,L_{\rm [O \,II]}}=0.3$
dex.  Note that Equation (4) is an empirical relation between observed
quantities.  It does not include dust extinction corrections.  The
inferred $L_{\rm [O\,II]}$ can therefore be directly compared to what
is observed in the stacked SDSS QSO spectra.  For a galaxy of $M_B$,
$L_{\rm [O\,II]}$ is drawn from a Gaussian distribution about the mean
$M_B$-$L_{\rm [O\,II]}$ relation with a 1-$\sigma$ width of 0.3 dex in
the $\log\,L_{\rm [O\,II]}$ space.

To examine whether or not our mock galaxy sample is representative of
the field galaxy population, we present the [O\,II] luminosity
distribution of our mock galaxy sample in Figure 2 along with the
observed [O\,II] luminosity function of $z\sim 1$ galaxies from
different deep surveys (Takahashi \etal\ 2007; Zhu \etal\ 2009).  All
the empirical measurements have been converted to have the same
cosmological parameters adopted in our analysis.  Figure 2
demonstrates that we reproduce the observed [O\,II] luminosity
function with the mock galaxy sample for the luminous galaxy
population.  At the faint-end, the observations suffer from survey
incompletness (e.g.\ Zhu \etal\ 2009) and therefore the observed space
density represents a lower limit to the underlying faint galaxy
population.

\begin{figure}
\includegraphics[scale=0.45]{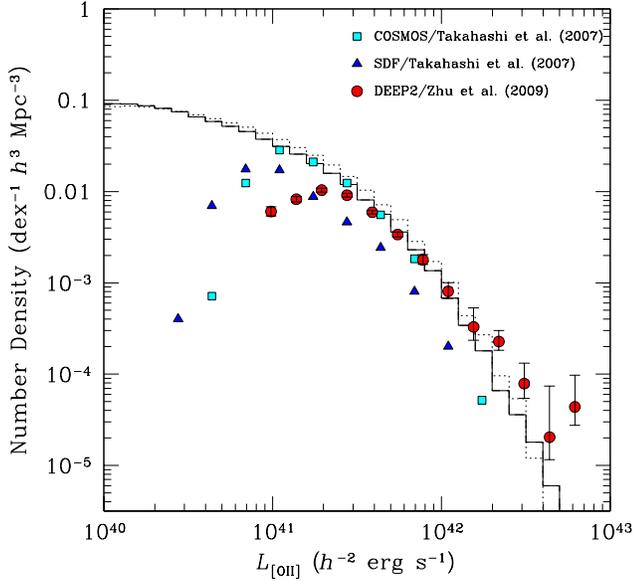}
\caption{The [O\,II] luminosity distribution of the mock galaxy sample
  described in Section 2.  The histograms display the simulated data
  for different faint-end slope of $\alpha=-1.2$ (dotted histogram)
  $\alpha=-1.3$ (solid histogram), and $\alpha=-1.4$ (dashed
  histogram).  The [O\,II] luminosity function measured for $z\sim 1$
  galaxies in DEEP2 (closed circles; Zhu \etal\ 2009) and for $z\sim
  1.2$ galaxies in the COSMOS field (closed squares; Takahashi \etal\
  2007) and the SDF (closed triangles; Takahashi \etal\ 2007) are also
  shown for comparison.}
\end{figure}

Next, we model each galaxy in the mock catalog as a round disk and
randomly place the galaxy within $\approx 10''$ of a QSO sightline and
at redshift between $z=0.4$ and $z=1.4$.  We compute the corresponding
physical projected distance of the galaxy based on the redshift and
angular distance to the QSO.  For each disk, we also assign a random
inclination angle with respect to the observer and a random position
angle of the major axis of the inclined disk with respect to the line
connecting the galaxy and the QSO.

%inclination angle, which is the angle of the normal vector of the
%galaxy with the plane of observation. The position angle was randomly distributed between
%$0^{\circ}$ and $180^{\circ}$. The inclination angle of the galaxy
%followed a sine distribution, which is due to the effects of projecting a disk in three dimensional space onto a two dimensional plane, between $0^{\circ}$ and
%$90^{\circ}$,\,with $0^{\circ}$ representing a face on observation and
%$90^{\circ}$ representing an edge on observation. 

To determine the fraction of the luminous disk that overlaps the SDSS
fiber centered at the QSO, we adopt the size of each disk galaxy
$R_{90}$ and its relative distance and orientation to the QSO.  We
de-project the inclined disk and determine the galactocentric distance
$R$ of each point ($x$, $y$) in the disk according to
\begin{equation}
  R^2(x,y)=r^2(x,y)\,\left[1+\cos^2{(\theta+\alpha)}\,\tan^2{i} \right]
\end{equation}
where $r$ is the projected radius on the plane of the sky, $\alpha$ is
the position angle of the major axis of the inclined disk, $\theta$ is
the azimuthal angle of point ($x$, $y$) from the major axis, and $i$
is the inclination angle of the disk.  The expected [O\,II] emission
from the galaxy in the QSO spectral data is then computed by
integrating all the light within $R_{90}$ of the disk that falls in
the $3''$ diameter fiber.  We consider two different surface
brightness profiles: (1) a flat, top-hat profile and (2) an
exponential profile for this calculation.  If the luminous disk does
not overlap the $3''$ diameter fiber, then we set $L_{\rm [O\,II]}=0$.
Finally, we divide the computed $L_{\rm [O\,II]}$ by the physical area
of a $3''$ diameter fiber at the redshift of the mock galaxy in order
to calculate its [O\,II] luminosity surface density $\sum_{L_{\rm
    [O\,II]}}$ that would be recorded in an SDSS QSO fiber.

\subsection{The Mock Mg\,II Absorber Catalog}

The mock Mg\,II catalog is formed by calculating the expected Mg\,II
absorption strength in the spectrum of the background QSO for every
galaxy in the mock galaxy catalog.  To determine the associated Mg\,II
absorption strength of a galaxy in the QSO spectrum, we adopt the
uniform gaseous halo model of Chen \etal\ (2010a).  Under this model,
the Mg\,II absorber strength for a galaxy at projected distance $\rho$
is characterized by a mean relation of
\begin{equation}
W_r(2796)=\frac{\ W_0}{a_h\,\sqrt{\rho^2/a_h^2+1}}\,\tan^{-1}{\sqrt{\frac{R_{gas}^2-\rho^2}{a_h^2+\rho^2}}}
\end{equation}
and dispersion $\sigma_{\log\,W_r}=0.233$, where $R_{\rm gas}$ is the
gaseous radius of the Mg\,II halo, $a_h$ is the core radius and is
$a_h=0.2\,R_{gas}$ and $\log\,W_0=1.24\pm0.03$.  The gaseous radius is
determined based on the galaxy $B$-band luminosity following a
power-law model,
\begin{equation}
\frac{R_{\rm gas}}{R_{\rm gas_*}}=\left(\frac{L_B}{L_{B_*}}\right)^{\beta},
\end{equation} 
for which Chen \etal\ (2010a) found a best-fit characteristic radius
of $R_{\rm gas_*}=75\ h^{-1}$ kpc for an $L_*$ galaxy and a scaling
index of $\beta=0.35$.  The scaling relation between gaseous radius
and galaxy $B$-band luminosity is understood as more massive galaxies
are surrounded by more extended halos (e.g.\ Tinker \& Chen 2008; Chen
\etal\ 2010b).  In addition to the spatial profile of the extended
Mg\,II gas, Chen \etal\ (2010a) also measured a high gas covering
fraction $\kappa$ within $R_{\rm gas}$.  Specifically, they found a
mean covering fraction of $\kappa_{0.3}\approx 100$\% for absorbers of
$\ewr\ge 0.3 $ \AA\ at $\rho \apl 0.4\,R_{\rm gas}$ and
$\kappa_{0.3}\approx 70$\% at $\rho\le R_{\rm gas}$.  The gas covering
fraction increases with decreasing \ewr\ threshold and decreases with
$\rho$ (see Figure 10 in Chen \etal\ 2010a).

For each simulated galaxy, we compute the expected \ewr\ by randomly
sampling the mean relation Equations (6) within the observed scatter
$\sigma_{\log\,W_r}$.  For simulated dwarf galaxies with $R_{\rm
  gas}<\rho$, we set $\ewr=0$.  Including the large scatter
($\sigma_{\log\,W_r}$) in the computation of \ewr\ allows the
possibility of galaxies at small $\rho$ producing Mg\,II absorbers
that are much weaker than the mean.  Although every galaxy at $\rho <
R_{\rm gas}$ in the mock sample is expected to produce a Mg\,II
absorber, the gas covering fraction measured at a given \ewr\
threshold is not 100\%.

For a mock sample of 200,000 galaxies, the procedure described above
produces a mock sample of $\sim 23,000$ Mg\,II absorbers of $\ewr\ge
0.3$ \AA\ at $z=0.4-1.4$.  The frequency distribution function of the
mock Mg\,II absorber sample is presented in Figure 3.  For comparison,
we also include in Figure 3 the model from Prochter \etal\ (2006) that
best describe the incidence of Mg\,II absorbers identified at
$z=0.4-2.3$ in their blind survey.  Aside from the offset in the
normalization, Figure 3 shows that the frequency distribution function
of the mock Mg\,II absorber sample follows the shape of the best-fit
model of Prochter \etal\ (2006).  The difference in the incidence of
mock Mg\,II absorbers (up to 20\%) for different adopted faint-end
slope of the galaxy luminosity function is understood by the
relatively steep dependence of halo gas cross section on galaxy
luminosity, $\sigma_{\rm gas}=\pi\,R_{\rm gas}^2\propto L_B^{0.7}$
from Equation (7).  Adopting known galaxy luminosity functions at
$z<1$ from Faber \etal\ (2007) and the scaling relation of Equation
(7), we expect to find (following Equation 8 below) a number density
of $n(z) \approx 0.8\ (0.2)$ per line of sight for Mg\,II absorbers of
$\ewr > 0.3\ (1.0)$ \AA.  Our model therefore well reproduces the
observe absorber statistics at $z<1$ (e.g.\ Nestor \etal\ 2005;
Prochter \etal\ 2006).

\begin{figure}
\includegraphics[scale=0.45]{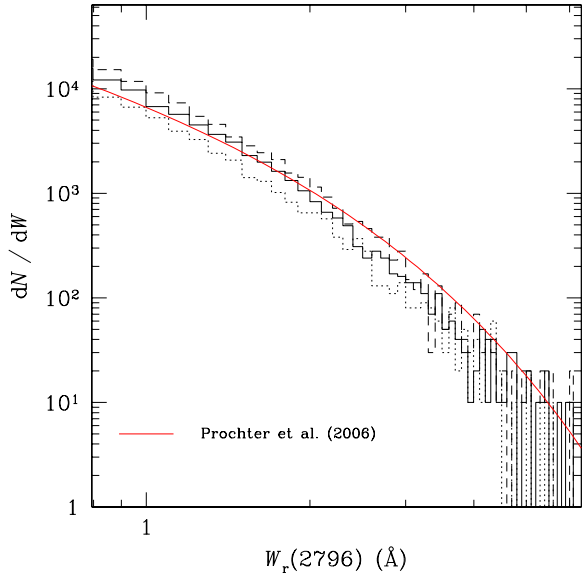}
\caption{The frequency distribution function of the mock Mg\,II
  absorber sample. As in Figure 1, the dotted, solid, and dashed
  histograms represent mock absorber samples from simulated galaxies
  of faint-end slopes of $\alpha=-1.2$, $-1.3$, and $-1.4$,
  respectively.  For comparison, we also include the observed absorber
  frequency distribution function from Prochter \etal\ (2006) for
  Mg\,II absorbers found at $z=0.4-2.3$.}
\end{figure}

%\clearpage
%%%%%%%%%%%%%%%%%%%%%%%%%%%%%
%%%%%%%%%%%%%%%%%%%%%%%%%%%%%
%        Analysis           %
%%%%%%%%%%%%%%%%%%%%%%%%%%%%%
%%%%%%%%%%%%%%%%%%%%%%%%%%%%%

%%%%%%%%%%%%%%%%%%%%%%%%%%%%% Incomplete
\section{Analysis}

Using the mock galaxy and Mg\,II absorber catalogs, we proceed to
examine the relation between \ewr\ and $L_{\rm [O\,II]}$ of Mg\,II
absorbers.  For each Mg\,II absorber in the mock catalog, we know from
our simulation the location and luminosity of the absorbing galaxy.
We have also calculated according to the procedures described in
Section 2.1 the fraction of galaxy light that would be recorded in the
SDSS QSO fibers.  We can therefore directly compare our simulation
data with the observations of M11.

We present in Figure 4 the distribution of \ewr\ and $\sum_{L_{\rm
    [O\,II]}}$ of the mock Mg\,II absorber sample.  The top panel
shows the \ewr--$\sum_{L_{\rm [O\,II]}}$ distribution of individual
Mg\,II absorbers, including detections (dots) and non-detections
(indicated by the arrows).  The bottom panel shows the fraction of
Mg\,II absorbers for which the absorbing galaxies are not expected to
overlap the fiber and therefore have $\sum_{L_{\rm [O\,II]}}=0$ in the
QSO spectra.  It is clear that a growing fraction of galaxy light is
missed in the QSO spectra for absorbers of decreasing strength due to
the \ewr\ vs.\ $\rho$ anti-correlation (e.g.\ Chen \etal\ 2010).

To reproduce the observations in SDSS QSO spectra, we divide the mock
catalog of $23,000$ Mg\,II absorbers into subsamples according to
their absorption strengths.  Following the proceduce described in M11,
who formed a median QSO spectrum at the rest-frame of the Mg\,II
absorbers and measured the associated [O\,II] emission line flux, we
compute the median value of [O\,II] luminosity surface density
$\langle\sum_{L_{\rm [O\,II]}}\rangle_{\rm med}$ for all Mg\,II
absorbers in each \ewr\ bin, including those with $\sum{L_{\rm
    [O\,II]}}=0$.  To estimate the scatter, we repeat the Monte Carlo
simulation 100 times to generate 100 mock samples of galaxies and
Mg\,II absorbers.  We measure the 1-$\sigma$ dispersion in
$\langle\sum_{L_{\rm [O\,II]}}\rangle_{\rm med}$ among the 100 mock
samples.  The stars and the associated errorbars in Figure 4 represent
the computed $\langle\sum_{L_{\rm [O\,II]}}\rangle_{\rm med}$ and the
associated dispersion in each \ewr\ bin.

For comparison, the best-fit power-law model of the observed
\ewr--$\langle\sum_{L_{\rm [O\,II]}}\rangle_{\rm med}$ correlation
(Equation 1) from M11 is included in Figure 4 as the dash-dotted line.
The M11 model is found to match well with the simulated data.
However, both fall below the locus defined by individual dots because
of non-detections quantified in the bottom panel.  We also calculate
the mean [O\,II] flux $\bar{\sum}_{L_{\rm [O\,II]}}$ expected in the
stacked SDSS QSO spectra, averaged over all Mg\,II absorbers in each
bin including non-detections.  The results are shown in solid circles
in Figure 4.

%Following M\'enard \etal\ (2011), who formed a composite
%QSO spectrum in the rest-frame of intervening Mg\,II absorbers by {\it
%  median-filtering} continuum-subtracted QSO spectra and searched for
%associated [O\,II] emission line in the composite spectrum.

\begin{figure}
\includegraphics[scale=0.45]{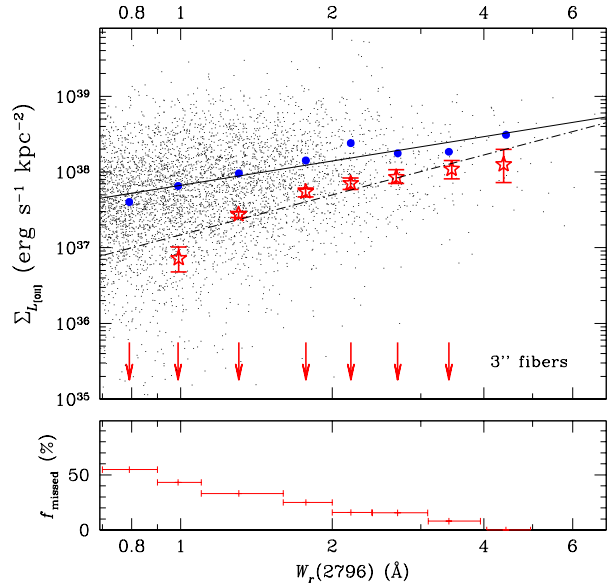}
\caption{The \ewr\ vs.\ $\sum_{L_{\rm [O\,II]}}$ distribution of the
  mock galaxy and absorber samples, in comparison to observations of
  M11.  The dots in the top panel represent individual Mg\,II
  absorbers with detectable emission fluxes in the QSO spectra from
  their absorbing galaxies.  Recall that these are calculated by
  dividing the expected [O\,II] line flux in the fiber with the
  corresponding physical area of the $3''$ fiber at the redshift of
  the absorber.  The arrows at the bottom row indicate the presence of
  absorbing galaxies with $\sum_{L_{\rm [O\,II]}}=0$ (non-detections),
  because their optical disks do not overlap with the fiber.  For
  completeness, the bottom panel shows the fraction of Mg\,II
  absorbers that have $\sum_{L_{\rm [O\,II]}}=0$.  The (red) stars in
  the top panel represent the {\it median} values of $\sum_{L_{\rm
      [O\,II]}}$ for {\it all} Mg\,II absorbers (including detections
  and non-detections) in each \ewr\ bin with the 1-$\sigma$ dispersion
  indicated by the errorbars.  The dash-dotted line shows the best-fit
  power-law model (Equation 1) of the observed
  \ewr--$\langle\sum_{L_{\rm [O\,II]}}\rangle_{\rm med}$ correlation
  from M11.  The solid (blue) circles represent the {\it mean}
  $\sum_{L_{\rm [O\,II]}}$ averaged over all Mg\,II absorbers in each
  $\ewr$ bin.  Note that the stars fall below the locus defined by
  individual dots because of non-detections quantified in the top
  panel.}
\end{figure}

% We observe that the binned data points representing the observable
% data follow the relation from \cite{menard}. Applying a least-fit
% square regression to the data generated by the 100 binning trials
% yields values of $A=1.76$ and $b=37.13$ for the constants in
% equation (\ref{correlation}).  Comparing these values to the values
% of $A=1.75$ \& $b=37.17$ from \cite{menard} shows that our
% simulation was able to reproduce the observed correlation.

%%%%%%%%%%%%%%%%%%%%%%%%%%%%%
%%%%%%%%%%%%%%%%%%%%%%%%%%%%%
%     	  DISCUSSION        %
%%%%%%%%%%%%%%%%%%%%%%%%%%%%%
%%%%%%%%%%%%%%%%%%%%%%%%%%%%%

\section{Discussion and Summary}

We have carried out a Monte Carlo study to investigate the effect of
differential aperture loss of extended emission from intervening
galaxies in SDSS fibers.  We generate a mock galaxy sample based on
known empirical relations that describe the luminosity and size
distribution of the general galaxy population uncovered in deep
surveys.  The mock galaxy sample is accompanied by a mock Mg\,II
absorber sample that is generated based on a simple assumption that
extended gaseous halos are a common and generic feature of distant
galaxies and the gaseous extent scales proportionally with galaxy
mass.  For each absorber in the mock sample, the luminosity and
projected distance of the absorbing galaxy are known, allowing us to
make predictions for the observed relation between $\ewr$ and the
associated $L_{\rm [O\,II]}$, in SDSS QSO spectra.

Our study shows that combining the known \ewr\ vs.\ $\rho$
anti-correlation of Mg\,II--galaxy pairs (e.g.\ Chen \etal\ 2010a) and
differential fiber selection of Mg\,II absorbing galaxies (Figure 1)
reproduces the observed \ewr\ vs. $L_{\rm [O\,II]}$ correlation in the
SDSS QSO spectra without additional fine-tuning or scaling.  The
results of our study indicate that the observed \ewr\ vs. $L_{\rm
  [O\,II]}$ correlation of Mg\,II absorbers in SDSS data is likely due
to a differential fiber selection bias and does not provide a physical
understanding of the origin of the Mg\,II absorber population.

On the basis of the the observed \ewr\ vs. $L_{\rm [O\,II]}$
correlation, M11 further attempted to draw connections between
%also pointed out two empirical
%measurements that support their conclusion that Mg\,II absorbers
%originate in outflows.  First, the frequency distribution function and
%the [O\,II] luminosity function share similar shape and amplitude.
the number density evolution of Mg\,II absorbers $n(z)$ and the cosmic
star formation history of the universe as characterized by the
comoving [O\,II] luminosity density $\ell$.  Given a good agreement
between $n(z)$ and $\ell$, the authors argue that outflows are the
mechanism responsible for Mg\,II absorption.

The Monte Carlo simulation presented in Section 2 also allows us to
address the agreements in these measurements under a simple, generic
halo model.  It is clear from Figures 3 \& 4 show that both the
frequency distribution function of Mg\,II absorbers and the galaxy
[O\,II] luminosity function are well reproduced in our mock galaxy and
absorber samples with no preference in starburst systems.  Although
under the simple halo model Mg\,II absorbers do not directly probe
star-forming disks like high column density damped \lya\ absorbers
(e.g.\ Wild \etal\ 2007), they trace the halo gas of distant galaxies.
If more massive galaxies are surrounded by more extended Mg\,II
absorbing gas (Equation 7), then $n(z)$ is calculated according to
\begin{equation}
n(z) = \frac{c}{H_0} \frac{(1 + z)^2}{\sqrt{\Omega_{M}(1+z)^3+\Omega_{\Lambda}}} \int dL \ \Phi(L,z)\, \kappa_{\rm gas}\,\sigma_{\rm gas}
\end{equation}
where $c$ is the speed of light, $\Phi(L_B,z)$ is the galaxy
luminosity function, $\kappa_{\rm gas}$ is the incidence of extended
gas that is the product of halo gas covering fraction and the fraction
of galaxies with extended gaseous halos, and $\sigma_{\rm gas}(L)$ is
the cross section of the gaseous halo.  As mentioned in \S\ 2.2, the
mean covering fraction of Mg\,II absorbing gas is found to be high,
roughly 100\% for absorbers of $\ewr\ge 0.3 $ \AA\ at $\rho \apl
0.4\,R_{\rm gas}$ (Chen \etal\ 2010a).  Such high covering fraction is
consistent with the analysis presented in Kacprzak \etal\ (2008) for a
scaling index of $\beta=0.35$.

Combining Equations (4) and (7) yields
\begin{equation}
\sigma_{\rm gas}(L) = \frac{\kappa_{\rm gas}\,(\pi\,R_{\rm gas}^2*)}{L_{\rm [O\,II]}^{0.93}*}\,L_{\rm [O\,II]}^{0.93} \approx  \frac{\kappa_{\rm gas}\,(\pi\,R_{\rm gas}^2*)}{L_{\rm [O\,II]}*}\,L_{\rm  [O\,II]}.
\end{equation}
Substituting Equation (9) into Equation (8) leads to
\begin{eqnarray}
n(z)&\approx&\frac{\kappa_{\rm gas}\,(\pi\,R_{\rm gas}^2*)}{\varepsilon(z)\,L_{\rm [O\,II]}*}\int L\,\Phi(L,z)\,d\,L \nonumber \\
    & = & \frac{\kappa_{\rm gas}\,(\pi\,R_{\rm gas}^2*)}{\varepsilon(z)\,L_{\rm [O\,II]}*}\,\ell_{\rm [O\,II]}(z),
\end{eqnarray}
where $\varepsilon(z)$ is defined so that
$1/\varepsilon(z)=(c/H_0)\,(1 +
z)^2/\sqrt{\Omega_{M}(1+z)^3+\Omega_{\Lambda}}$.  The same redshift
dependent factor is defined in Equation (7) of M11.  Equation (10)
shows that the number density evolution of absorbers naturally follows
the comoving [O\,II] luminosity density, $\ell_{\rm [O\,II]}(z)$, if
extended gaseous halos are a common and generic feature of field
galaxies.  One can therefore apply $n(z)$ for constraining the cosmic
star formation history, or apply known co-moving luminosity density
for constraining the fraction of absorbers originating in galactic
halos (Chen \etal\ 2000).

In practice, as illustrated in M11 the SDSS $3''$ fibers define a
survey volume of [O\,II] emitting galaxies.  The comoving [O\,II]
luminosity density, $\ell_{\rm [O\,II]}(z)$, can be estimated
according to
\begin{eqnarray}
\ell_{\rm [O\,II]}(z) &\equiv& \frac{\Delta\,L_{\rm [O\,II]}}{\Delta\,V_c} \nonumber \\
        &=& \frac{1}{(1+z)^2}\frac{1}{D_A^2(z)\,d\,\Omega}\frac{\Delta\,L_{\rm [O\,II]}}{\Delta\,N}\frac{\Delta\,N}{\Delta\,r} \nonumber \\
        &=& \frac{\Delta\,z}{(1+z)^2\,\Delta\,r}\frac{\bar{L}_{\rm [O\,II]}}{D_A^2(z)\,d\,\Omega}\frac{\Delta\,N}{\Delta\,z} \nonumber \\
        &=& \varepsilon(z)\,\bar{\sum}_{L_{\rm [O\,II]}}\,n(z),
\end{eqnarray}
where $\Delta\,L_{\rm [O\,II]}$ is the total [O\,II] luminosity
observed in a coadded QSO spectrum, $\Delta\,N$ is the total number of
Mg\,II absorbers over the survey redshift pathlength $\Delta\,z$, and
$\bar{L}_{\rm [O\,II]}$ is the mean [O\,II] luminosity averaged over
the entire galaxy (absorber) population.  It is clear that Equation
(11) resembles Equation (10) and that $\ell_{\rm [O\,II]}(z)$ can be
estimated based on the product of $n(z)$ and the mean [O\,II]
luminosity surface density averaged over {\it all} Mg\,II absorbers.

We present in Figure 4 the mean [O\,II] luminosity surface density
$\bar{\sum}_{L_{\rm [O\,II]}}$ measured in the $3''$ SDSS fibers and
averaged over the number Mg\,II absorbers in each \ewr\ bin (solid
circles).  The mean values can be characterized by
\begin{equation}
\bar{\sum}\,_{L_{\rm [O\,II]}}=a_{3''}\left[\frac{\ewr}{1\,{\rm \AA}}\right]^{b_{3''}}
\end{equation}
where $a_{3''}=6.6\times 10^{37}\,{\rm erg}\,{\rm s}^{-1}\,{\rm
  kpc}^2$ and $b_{3''}=1.1$ (solid line in Figure 4).  This
correlation applies to a $3''$ survey cylinder at $z=0.4-1.4$.  Note
that because of the extreme bimodal distribution of $\sum_{L_{\rm
    [O\,II]}}$ in which a large fraction of absorbers have
$\sum_{L_{\rm [O\,II]}}=0$, the median value $\langle\sum_{L_{\rm
    [O\,II]}}\rangle_{\rm med}$ is expected to be significantly
smaller than the mean $\bar{\sum}_{L_{\rm [O\,II]}}$.

While Equation (11) provides a useful tool for estimating the
co-moving star formation rate density $\dot{\rho}_*$ based on the
observations of Mg\,II absorbers in SDSS QSO spectra (as shown
cleverly in M11), we note that the observed strong \ewr\ vs.\
$\sum_{L_{\rm [O\,II]}}$ correlation in SDSS data is shaped primarily
by a differential fiber loss of the observed [O\,II] flux and
therefore unphysical.  To illustrate the fiber selection bias in
driving the apparent correlation in Figure 4, we repeat the Monte
Carlo analysis described in \S\ 2 for $15''$ diameter fibers.  This is
five times the aperture size of SDSS spectra.  At $z\approx 1$, such a
large aperture covers an area of roughly $40\ h^{-1}$ kpc projected
radius from QSO lines of sight, sufficient to cover the majority of
the light from galaxies producing strong Mg\,II absorbers ($\ewr\apg
1$ \AA).  Figure 5 shows the distribution of \ewr\ and $\sum_{L_{\rm
    [O\,II]}}$ measured over $15''$ diameter fibers for the mock
Mg\,II absorber sample.  The mean [O\,II] flux $\bar{\sum}_{L_{\rm
    [O\,II]}}$ share a similar slope as the median [O\,II] flux
$\langle\sum_{L_{\rm [O\,II]}}\rangle_{\rm med}$ within the $15''$
diameter aperture and the slope is consistent with $b_{15''}=0$ (solid
line in Figure 5).

\begin{figure}
\includegraphics[scale=0.45]{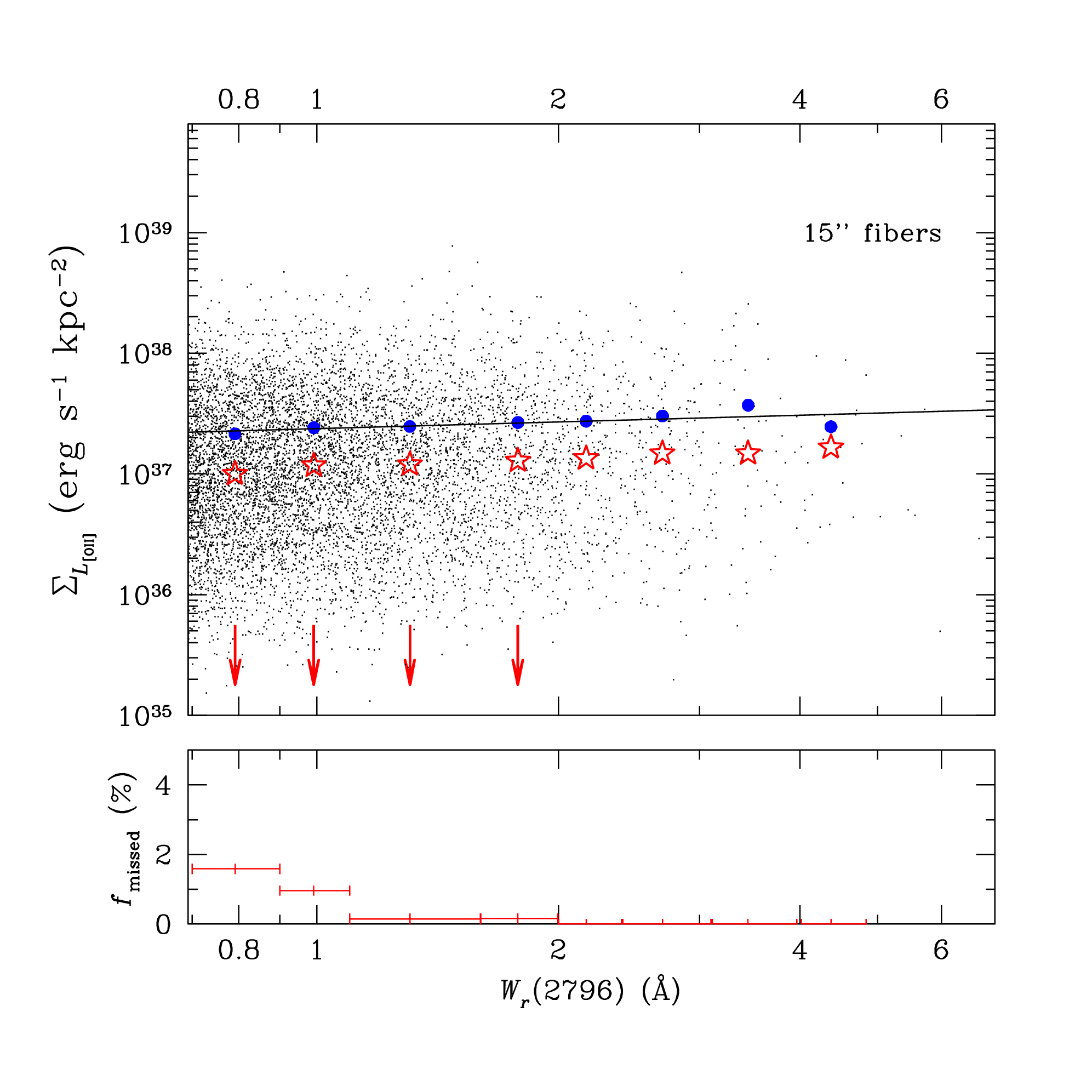}
\caption{The same as Figure 4, but assuming $15''$ fibers in the Monte
  Carlo simulation described in \S\ 2.  Both $\langle\sum_{L_{\rm
      [O\,II]}}\rangle_{\rm med}$ (red stars) and $\bar{\sum}_{L_{\rm
      [O\,II]}}$ (solid blue circles) show a lack of correlation with
  \ewr.  The bottom panel shows that with $5\times$ larger fibers,
  differential aperture loss has minimal impact on the survey of
  Mg\,II selected galaxies along QSO lines of sight. }
\end{figure}

In summary, we have demonstrated that (1) because of the observed
$\ewr$ vs. $\rho$ anti-correlation (e.g.\ Chen \etal\ 2010a), galaxies
at larger projected distances produce on average weaker Mg\,II
absorbers and weaker (or zero) $L_{\rm [O\,II]}$ in SDSS QSO spectra
and that (2) because of the extreme bimodal distribution of
$\sum_{L_{\rm [O\,II]}}$ in which a large fraction of absorbers have
$\sum_{L_{\rm [O\,II]}}=0$, the median value $\langle\sum_{L_{\rm
    [O\,II]}}\rangle_{\rm med}$ can be significantly smaller than the
mean $\bar{\sum}_{L_{\rm [O\,II]}}$.  Together these effects strenthen
the apparent $\ewr$--$\langle\sum_{L_{\rm [O\,II]}}\rangle_{\rm med}$
correlation of M11.

Consequently, empirical correlations between star-forming properties
of galaxies and statistical properties of QSO absorbers do not provide
the unambiguous evidence necessary to discriminate between infalling
clouds and outflows as the mechanism for producing Mg\,II absorbing
clouds at $\rho>50$ kpc of a galaxy.  While outflows are a natural
product of starbursts, gas accretion provides the fuels for star
formation in galaxies.  As discussed by previous authors, mass is
likely a more fundamental factor that determines the properties of
gaseous halos around galaxies (e.g.\ Steidel \etal\ 1994; Ledoux
\etal\ 2006; Chen \etal\ 2010b) and more massive galaxies can support
more extended gaseous halos.  Given that models that do not require
outflows can also reproduce the empirical correlations between
absorber and star formation properties, we caution drawing conclusions
in favor of outflows based on simple correlations between absorber and
star formation properties.

\section*{Acknowledgments}

We thank Jean-Ren\'e Gauthier, Michael Rauch, Rob Simcoe, Ben Weiner,
Vivienne Wild, and Art Wolfe for helpful discussions and comments.
G.L. acknowledges support from the Physics REU program at the
University of Chicago.
% H.-W.C. acknowledges support from an NSF grant
%AST-0607510.

%\clearpage\\] 

%\label{lastpage}

\end{document}